\documentclass[pra,aps,twocolumn]{revtex4}
\usepackage{bm,graphicx,amsmath} \usepackage{bbm}

\newcommand{\br}[1]{\langle #1|}
\newcommand{\ke}[1]{|#1\rangle}

\newcommand{\al}[1]{^{(#1)}}
\newcommand{\da}{^\dagger}

\newcommand{\pt}[1]{\left( #1 \right)}
\newcommand{\pq}[1]{\left[ #1 \right]}
\newcommand{\pg}[1]{\left\{ #1 \right\}}

\newcommand{\lpq}[1]{\left[ #1 \right.}

\newcommand{\rpq}[1]{\left. #1 \right]}

\begin{document}

\title{Entanglement of distant atoms by
projective measurement:\\ The role of detection efficiency}

\author{Stefano Zippilli,$^{1}$, Georgina A. Olivares-Renter{\'i}a,$^{1,2}$ and
Giovanna Morigi$^{1}$}
\affiliation{
$^{1}$Departament de F\'{i}sica, Universitat Aut\`{o}noma de Barcelona, E-08193 Bellaterra, Spain.\\
$^{2}$Center for Quantum Optics and Quantum Information, Departamento de F\'{i}sica,
Universidad de Concepci\'{o}n, Casilla 160-C, Concepci{\'o}n, Chile.}
\author{Carsten Schuck,$^{3}$ Felix Rohde,$^{3}$ and
J\"urgen Eschner$^{3}$} \affiliation{ $^{3}$ ICFO-Institut de Ci\`{e}ncies
Fot\`{o}niques, E-08860 Castelldefels, Barcelona, Spain. } \date{\today}

\begin{abstract}
We assess proposals for entangling two distant atoms by measurement of emitted photons,
analyzing how their performance depends on the photon detection efficiency. We consider
schemes based on measurement of one or two photons and compare them in terms of the
probability to obtain the detection event and of the conditional fidelity with which the
desired entangled state is created. Based on an unravelling of the master equation, we
quantify the parameter regimes in which one or the other scheme is more efficient,
including the possible combination of the one-photon scheme with state purification. In
general, protocols based on one-photon detection are more efficient in set-ups
characterized by low photon detection efficiency, while at larger values two-photon
protocols are preferable. We give numerical examples based on current experiments.
\end{abstract}

\maketitle

\section{Introduction}

Quantum networks based on atom-photon interfaces require full control of the atom-photon
interactions and correlations~\cite{ZollerRoadmap}. A major issue, related to the
realization of quantum repeaters~\cite{ZollerRoadmap,Briegel98}, is the entanglement of
distant nodes. Existing proposals for establishing entanglement by deterministic
interactions are usually very demanding on the technological side~\cite{Cirac97,Kraus04}.
Present experimental effort is therefore partially focussed on the realization of
entanglement of the internal degrees of freedom of distant atoms by projective
measurement through photodetection~\cite{Cabrillo,Browne,FengDuanSimon}. These
protocols are essentially based on first entangling atoms with their emitted photons, and
then performing entanglement swapping from the photons onto the atoms by photodetection
in a Bell measurement setup. In this context, several experiments have analyzed the
coherence properties of photons emitted by distant atoms \cite{Eichmann1993, Eschner2001,
Walther04, Kimble04, Grangier05, Grangier06, Kuhn2, Rempe07, Kimble08}. Atom-photon
entanglement was demonstrated in Ref.~\cite{Monroe04,Weinfurter06}. More recently,
in~\cite{Maunz} the protocol described in~\cite{FengDuanSimon} was applied, thereby
realizing entanglement of two distant atoms.

These investigations with single atomic systems are complementary to experimental studies
which aim at realizing photonic interfaces with atomic ensembles. In this context,
entanglement of the collective spins of distant atomic ensembles by photodetection was
demonstrated in the experiments of Refs.~\cite{Kimble-Ensemble, Kuzmich-Ensemble, Polzik-Teleportation,Schmiedmayer}.
Theoretical investigations are also studying interface implementations with different
types of systems, such as solid-state based quantum bits~\cite{Lukin}. More complex
constructions have been recently theoretically discussed, which in principle allow for
generating many-particle entangled states~\cite{LamataCirac,Solano}.

An important issue in protocols based on quantum state projection by photodetection is
the role of the detection efficiency: detector quantum efficiency and finite photon
collection efficiency of the optical setup constitute intrinsic limits to the fidelity of
the produced entanglement, as well as to the rate at which a protocol can be successfully
realized.

In this work we focus on protocols for entangling two distant individual atomic systems
\cite{Cabrillo,Browne,FengDuanSimon} and analyze quantitatively how the photon detection
efficiency affects their performance. We use two criteria to assess the protocols, the
success probability with which their execution produces the required detection event, and
the conditional fidelity, corresponding to the probability of finding the target
entangled state after that event. We calculate these criteria from an unravelling of the
quantum optical master equation describing the respective protocols. In their comparison,
we find that schemes based on detection of two photons exhibit larger fidelity, while
schemes requiring detection of a single photon have usually larger success probability.
The latter ones therefore lend themselves to applying purification protocols. In
particular, we study application of the purification protocol proposed in
Ref.~\cite{Deutsch} to the entanglement scheme of Ref.~\cite{Cabrillo} and identify the
regimes of detection efficiency in which proposals based on single-photon detection are
more efficient than proposals requiring coincidence detection of two photons. In our
treatment we do not consider dark counts in the photodetection; their effect, as well as
the one of imperfections of the optical setup, have been analyzed in
Ref.~\cite{MonroeJOSAB}.

This article is organized as follows. In Sec.~\ref{Sec:1} we summarize the basic features
of the considered protocols and introduce the criteria we apply in order to determine
their efficiency. In Sec.~\ref{Sec:2} we introduce the theoretical tools we use for
calculating these criteria and which account explicitly for the detector efficiency. In
Sec.~\ref{Sec:3} we analyze and compare the efficiency of the various entanglement
protocols. Experimental considerations are reported in Sec.~\ref{Sec:4}, before we
conclude in Sec.~\ref{Sec:5}. In the appendix we report some additional details.

\section{Entanglement schemes}
\label{Sec:1}

In this section we summarize the basic features of recently proposed
protocols~\cite{Cabrillo,Browne,FengDuanSimon} which exploit atom-photon
correlations in order to achieve entanglement between the internal states of distant
atoms. The discussion is focussed on entanglement schemes, but teleportation generally
follows analogous procedures (see, for instance, Ref.~\cite{Bose}). In all the considered proposals, entanglement of the atomic
internal states is achieved by interference of photons emitted from distant atoms and
subsequent photo-detection projecting the atoms into a Bell state. These protocols are
probabilistic in the sense that the expected result is conditioned on the successful
detection of a photon.

In order to characterise the schemes we will apply two main criteria, the success
probability, $P_{\rm suc}$, and the conditional fidelity, $F$. The success probability of
a protocol measures which fraction of its executions leads to the desired photon
detection,
\begin{equation}
\label{Psucs} P_{\rm suc}= \frac {\rm \sharp~detection~events} {\rm
\sharp~scheme~executions}~.
\end{equation}
After a successful detection event, the conditional fidelity measures with which
probability the density matrix of the system, $\rho_M$, matches the density matrix of the
desired target state, $\rho_T$,
\begin{equation}
\label{fidelity} F={\rm Tr}\pg{\rho_T \rho_M}~.
\end{equation}
We will evaluate the state after the measurement, $\rho_M$, using an unraveling of the
master equation, where the detection event is explicitly introduced as an operator
determining the system evolution. This formalism is developed in the next section. A more
general (although less practical) quantity providing an overall comparison of the schemes
is the average fidelity, $\overline{F}$, which we define as
\begin{equation}
\label{F:av} \overline{F}=P_{suc}F~,
\end{equation}
and which corresponds to the total probability to obtain the desired state in one
execution of the entanglement protocol. We will apply these efficiency criteria
qualitatively when we present the various schemes in the reminder of this section; their
calculation and a detailed evaluation will be presented in later sections.

Since all of the considered protocols are conditioned on the detection of photons, the
fidelity and the success probability will depend on the efficiency, $\eta$, with which
these photons are detected, as well as on the probability, $p$, with which they are
emitted. Assuming that all optical elements perform perfect unitary local operation, and
that the initial atomic state can be prepared with unit fidelity, the detection
efficiency is determined by the finite photon-collection efficiency of the optical
apparatus, $\chi$, {\it i.e.}, the probability that an emitted photon reaches the detector, and
by the quantum efficiency of the detector itself, $\eta_d$, through
\begin{equation}
\label{eta} \eta=\chi\eta_d~.
\end{equation}

The basic system of all considered proposals are two atoms with $\Lambda$-shaped level
configuration, confined at two distant positions in space. Information is encoded in the
(meta-)stable internal states $\ke{e}_j$ and $\ke{g}_j$ which are coupled to a
common excited state $\ke{r}_j$, where $j=1,2$ labels the atoms. Both atoms, which may be
placed in free space or coupled to a resonator field, are initially prepared in the
ground state $\ke{e}_j$. They are then either excited by a laser pulse to an intermediate
state which thereafter evolves during a given detection time interval, or they are
continuously laser driven during the detection time interval. We distinguish protocols
which rely on the detection of one or of two photons during this detection time interval.

\subsubsection*{Single-photon schemes} 
\begin{figure}[!th] \begin{center}
\includegraphics[width=9cm]{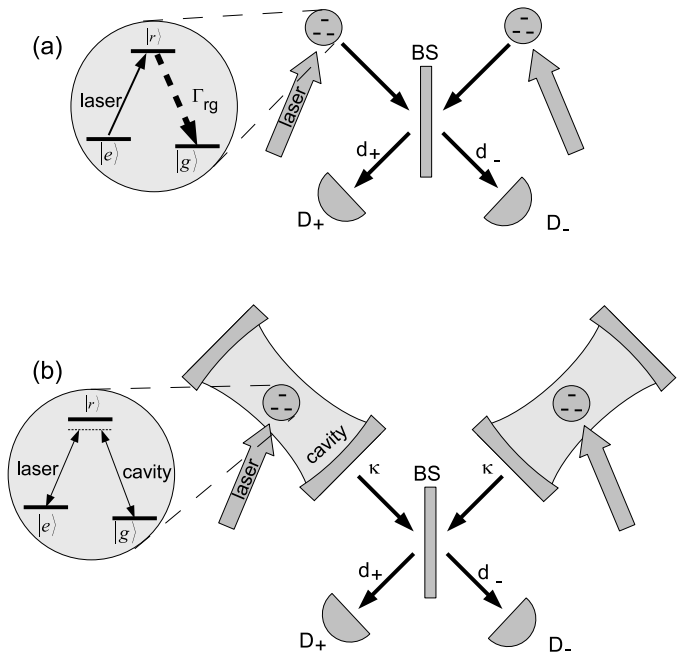}
\caption{Setups for entangling the internal degrees of freedom of two distant atoms by
detection of one photon. The transition $|e\rangle\to |r\rangle$ is driven by a laser.
Photons emitted on $|r\rangle\to |g\rangle$ are mixed on a 50:50 beam splitter (BS). A
click of a detector $D_{\pm}$ ideally projects the atoms in the Bell state of
Eq.~(\ref{Target}). In (a) the photon is emitted by spontaneous Raman scattering into
free space~\protect\cite{Cabrillo}, while in (b) it is emitted by coherent Raman
scattering into the mode of a cavity, which then decays into free
space~\protect\cite{Browne}.} \label{1:photon}
\end{center} \end{figure}

We first focus on the proposal by Cabrillo {\it et
al.}~\cite{Cabrillo}, based on the projective measurement of a single photon emitted from
either of two continuously excited atoms in free space. The typical setup is sketched in
Fig.~\ref{1:photon}(a).

The two atoms are initially prepared in the state $\ke{e}_1\ke{e}_2$, from where they are
weakly driven during the detection interval by a laser resonant with the
$\ke{e}_j\to\ke{r}_j$ transition. Once an atom has been excited to the state $\ke{r}_j$,
the protocol requires spontaneous emission along the transition $\ke{r}_j\to\ke{g}_j$
into one of the modes $\lambda$ of the free space electromagnetic field corresponding to
this transition. Denoting by $\varepsilon$ the probability amplitude associated with this
event, the atom-photon system is in the state
\begin{eqnarray} \label{State:Cabrillo}
|\Psi^I\rangle &=& \alpha^2 |e_1,e_2;0\rangle \nonumber \\
 &+& \varepsilon \alpha (|e_1,g_2;1_{\lambda}\rangle+|g_1,e_2;1_{\lambda'} \rangle) \\
 &+& \varepsilon^2 |g_1,g_2;1_{\lambda},1_{\lambda'}\rangle~, \nonumber
\end{eqnarray}
where $|\alpha|^2 + |\varepsilon|^2 =1$, and the states $|0\rangle$,
$|1_{\lambda}\rangle$ represent zero and one photon in the mode
$\lambda$~\cite{Footnote}.

To obtain an entangled state, one mixes the collected photons on a 50:50 beam splitter,
such that at its outputs one cannot determine which atom emitted the photon, {\it i.e.}, all
which-path information is erased. When additionally a stable phase relation between the
two scattering paths is established, {\it i.e.}, single-photon interference of the probability
amplitudes is warranted, then a click at the detector $D_{\pm}$ at one of the output
ports projects the atoms onto one of the Bell states
\begin{equation} \label{Target}
|\Psi^{\pm}\rangle= \frac{1}{\sqrt{2}}(|e_1,g_2\rangle\pm |e_2,g_1\rangle)~.
\end{equation}
In order to obtain this target state with high fidelity, one has to suppress the emission
of two photons, which would project the atoms into a product state. This may be achieved
by only very weakly exciting the atoms on the $\ke{e}_j\to\ke{r}_j$ transition, which
implies $|\varepsilon|^2\ll 1$ and hence entails low success probability. The success
probability is also limited by the finite solid angle within which photons emitted into
free space are collected, which leads to low detection efficiency. The latter problem can
be partially solved by confining each atom inside a resonator, such that the photons are
emitted preferably into the resonator modes which strongly couple to the atomic
transition $\ke{r}\to\ke{g}$~\cite{Browne}.

Instead of continuously exciting the atoms, one may apply a laser pulse to atoms placed
in cavities, like in the teleportation protocol proposed by Browne et al.~\cite{Browne},
whose setup is sketched in Fig.~\ref{1:photon}(b). During the initial pulse, the states
$\ke{e}_j$ and $\ke{g}_j$ are resonantly coupled via a stimulated Raman transition driven
by the laser and the cavity field, such that the atom-cavity system is prepared in the
state
\begin{eqnarray} \label{State:Browne}
|\Psi^I\rangle&=&\left(\alpha|e_1;0_1\rangle+\varepsilon|g_1;1_1\rangle\right)\\
              & &\times \left(\alpha|e_2;0_2\rangle+\varepsilon|g_2;1_2\rangle\right)\,,\nonumber
\end{eqnarray}
where $|n_j\rangle$ ($n=0,1$) corresponds to $n$ photons in the cavity $j$. During the
detection interval, starting after the preparatory laser pulse, photons may leak out from
the cavities via the mirrors of finite transmittance, and are then mixed on a 50:50 beam
splitter to erase any which-path information. Again, single photon interference and
subsequent detection of one photon projects the atoms onto the entangled target state
$|\Psi^{\pm}\rangle$ with a certain fidelity and success probability. Although the
detection efficiency may be high in this scheme due to the enhanced emission probability
into the cavity modes, the protocol is still limited by the finite probability that two
photons are emitted of which only one is detected during the measurement interval,
leaving the atoms in a product state.

\subsubsection*{Two-photon schemes} 
\begin{figure}[!th] \begin{center}
\includegraphics[width=9cm]{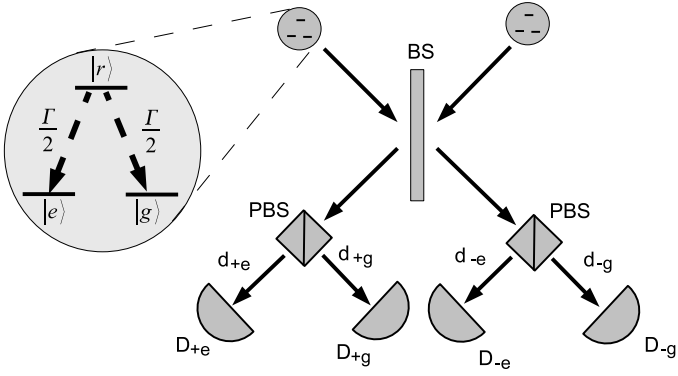}
\caption{Setup for entangling the internal states of two distant atoms by measurement of
two photons~\protect\cite{FengDuanSimon}. The atoms are prepared in the state
$|r_1,r_2\rangle$ and then decay spontaneously. The photon wavepackets overlap at a 50:50
beam splitter (BS). The detection apparatus at the output ports of the BS involves two
polarizing beam splitters (PBS) and four detectors. Coincident clicks at $D_{+,e}$ and
$D_{+,g}$ or $D_{-,e}$ and $D_{-,g}$ project the atoms into the state $|\Psi^+\rangle$.
The state $|\Psi^-\rangle$ is found by coincident clicks at $D_{+,e}$ and $D_{-,g}$ or
$D_{+,g}$ and $D_{-,e}$.}\label{3scheme}
\end{center} \end{figure}

The conceptually slightly
different approach discussed in~\cite{FengDuanSimon} relies on a (partial) Bell state
measurement of two photons, instead of single photon detection. In this protocol, the
atoms are located in free space and are initially excited to the state $|r_1,r_2\rangle$.
Each atom can decay along the two possible decay channels $|r\rangle \to |e\rangle$ and
$|r\rangle \to |g\rangle$, which are assumed to produce photons of orthogonal
polarisations, $\xi^{e}$ and $\xi^{g}$, at equal decay constants. The corresponding state
into which the atoms decay reads
\begin{eqnarray} \label{State:Simon}
|\Psi^I\rangle&=&\frac{1}{2}\left(|e_1;1_{\lambda_1,\xi^e}\rangle+|g_1;1_{\lambda_1,\xi^g}\rangle\right)\nonumber\\
              & &\times \left(|e_2;1_{\lambda_2,\xi^e}\rangle+|g_2;1_{\lambda_2,\xi^g}\rangle\right)\,,\nonumber
\end{eqnarray}
where the states $|1_{\lambda_j,\xi}\rangle$ represent one photon in the spatial mode of
emission corresponding to atom $j$ and polarization $\xi$~\cite{Footnote:Mode}. After the
emission each atom is maximally entangled with its corresponding photon
\cite{Monroe04,Weinfurter06}, and in case the two emitted photons have orthogonal
polarizations, the two atoms must be in different ground states. A partial Bell state
measurement performed on the two photons with the apparatus shown in Fig.~\ref{3scheme}
projects the atoms into the corresponding entangled state.

In contrast to the single-photon schemes, this protocol is based on two-photon
interference~\cite{HongOuMandel} of photon wavepackets overlapping on a beam splitter,
which makes it indistinguishable which atom decayed into which ground state.

The success probability for this scheme is low, as it relies on the detection of two
photons, both of which are recorded with finite probability, although the detection
efficiency could be enhanced by coupling the atoms to cavities. On the other hand, one
finds that the conditional fidelity in an ideal setup approaches unity for coincident
detection of two photons on certain detector combinations, see Fig.~\ref{3scheme}. This
protocol has been applied in the experiment of Ref.~\cite{Maunz}.

\section{The theoretical model}
\label{Sec:2}

In order to derive systematically the efficiency of the entanglement schemes under the
effect of finite detection probability, we use a master equation description for the
density matrix $\rho$ of photons and atoms. At time $t=0$ the system is in a given
initial state, which depends on the specific protocol as explained in the previous
section. Its dynamics are governed by the master equation
\begin{equation}
\label{masterEq0}
\frac{\partial}{\partial t}\rho=\pq{{\cal L}+J}\rho(t)\,,
\end{equation}
where ${\cal L}$ and $J$ are superoperators that describe the dynamics of photon emission
by the atomic system. They take the generic form
\begin{eqnarray}\label{L}
{\cal L}\rho&=&
-R\sum_{j=1,2}\sum_\xi\pq{{A^{(j)}_{\xi}}\da A^{(j)}_{\xi}\rho+\rho {A^{(j)}_{\xi}}\da A^{(j)}_{\xi}}\,,\nonumber\\
J\rho&=&2R\sum_{j=1,2}\sum_{\xi} A^{(j)}_{\xi}\rho
{A^{(j)}_{\xi}}\da\,,
\label{J0}
\end{eqnarray} where $2R$ is the photon emission
rate and the operators $A^{(j)}_{\xi}$ and ${A^{(j)}_{\xi}}\da$ describe respectively the
atomic dipole lowering and rising operators, when the photon is emitted by a spontaneous
decay of the atom $j=1,2$~\cite{Cabrillo,FengDuanSimon}. In protocols where the atoms are
inside a resonator~\cite{Bose,Browne}, they correspond instead to the annihilation and
creation operator of a cavity photon. The subscript $\xi$ labels the photon polarization,
which is relevant in the two-photon protocol~\cite{FengDuanSimon}.

In Eq.~(\ref{masterEq0}) the superoperator ${\cal L}$ describes the damping process,
while the superoperator $J$ is the so-called jump operator, describing the quantum-state
projection associated with the emission process. It can be decomposed into the sum
\begin{eqnarray}\label{J}
J\rho={\cal C}\rho +(1-\eta)J\rho\,,
\end{eqnarray}
where $\eta$ is the detection efficiency, defined in Eq.~(\ref{eta}), and ${\cal C}=\eta
J$ is the so-called click operator, describing quantum-state projection associated with a
click of the detection apparatus~\cite{Englert}. The second term on the Right-Hand-Side
(RHS) of Eq.~(\ref{J}) describes quantum-state projection associated with the photons
which are emitted but not detected. Using the unraveling of the master
equation~\cite{Carmichael,Englert,Dum} we write the solution of Eq.~(\ref{masterEq0}) as
\begin{eqnarray}\label{Eq:Unravel}
\rho(t)&=&U(t)\rho(0)+\int_0^t{\rm d}\tau U(t-\tau){\cal C} U(\tau)\rho(0)\\
& &+\int_0^t{\rm d}\tau\int_0^{\tau}{\rm d}\tau_1U(t-\tau){\cal
C}U(\tau-\tau_1){\cal C}U(\tau_1)\rho(0)\,,\nonumber
\end{eqnarray}
where the operator
\begin{equation}\label{U}
U(\tau)=\exp\left[({\cal L}+(1-\eta)J)\tau\right]
\end{equation}
gives the evolution conditioned on no click at the detector in the interval of time
$[0,\tau]$. Each term on the RHS of Eq.~(\ref{Eq:Unravel}) can be identified with
different detection  scenarios
over the time $t$ in which the experiment is run. The first term corresponds to the case
in which no click is recorded at the detector. The probability that this occurs is given
by
\begin{equation}\label{P:0}
P_0={\rm Tr}\{U(t)\rho(0)\}
\end{equation}
and the density matrix, conditioned on this event, is $\rho^{(0)} = U(t)\rho(0)/P_0$. The
second term corresponds to the case in which a click is recorded at time $\tau$ and no
further click until time $t$. The probability of this event is
\begin{equation}\label{P1}
P_1={\rm Tr}\pg{\int_0^t{\rm d}\tau U(t-\tau){\cal C}U(\tau)\rho(0)}
\end{equation}
and the corresponding density matrix is
\begin{eqnarray}
\rho_1(t)=\int_0^t{\rm d}\tau U(t-\tau){\cal C} U(\tau)\rho(0)/P_1\,.
\end{eqnarray}
Finally, the last term on the
RHS of Eq.~(\ref{Eq:Unravel}) describes events in which two photons are revealed at the
detector at the instants $\tau_1$ and $\tau>\tau_1$, which occur at probability
\begin{eqnarray}\label{P2}
&&P_2={\rm Tr}\pg{\int_0^t{\rm d}\tau\int_0^{\tau}{\rm d}\tau_1U(t-\tau){\cal C}U(\tau-\tau_1){\cal C}U(\tau_1)\rho(0)}\nonumber\\
\end{eqnarray}
and produce the corresponding density matrix.

Photo-detection occurs in the schemes at the outputs of a beam splitter whose two input
ports are the modes into which atoms 1 and 2 (or cavities 1 and 2) emit. We now express
the click operator ${\cal C}$ through the projectors of the corresponding von-Neumann
measurement. This corresponds to decomposing the click operator into a sum accounting for
the possible clicks one can record,
\begin{eqnarray}
{\cal C}\rho= \sum_\xi\pq{{\cal C}_{+\xi}+{\cal C}_{-\xi}}\rho\,,
\end{eqnarray}
where ${\cal C}_{\pm\xi}$ describes quantum-state projection by recording a photon at the
output port $D_{\pm,\xi}$. Denoting by $d_{\pm\xi}$ and $d_{\pm\xi} \da$ the operators
for the output field after the beam splitter, then
$${\cal C}_{\pm\xi}\rho=2R\eta d_{\pm\xi}\rho{d_{\pm\xi}}\da~,$$
with
\begin{eqnarray}
d_{\pm\xi}=\frac{1}{\sqrt{2}}\left(A^{(1)}_{\xi}\pm A^{(2)}_{\xi}\right)\,,
\end{eqnarray}
where we have used the input-output formalism to connect the output field to the system
operators~\cite{MilburnWalls} and omitted to write the input noise
terms~\cite{Footnote:Inputnoise}.

Let us now identify the density matrix of the system after a particular detection event which is to be used in the evaluation of the fidelity in Eq~(\ref{fidelity}).
We consider the case in which one
click is recorded in the time interval $[0,t]$ at the detector $D_{+,\xi}$, while no
click has been recorded otherwise at any detector. This occurs with probability
\begin{equation}
P_{+,\xi}(t)=\int_0^t{\rm d}\tau{\rm Tr}\{U(t-\tau){\cal C}_{+,\xi}U(\tau)\rho(0)\}
\end{equation}
and the system is projected into the state
\begin{equation}\label{rho+xi}
\rho_{+,\xi}(t)=\int_0^t{\rm d}\tau U(t-\tau){\cal C}_{+,\xi}U(\tau)\rho(0)/P_{+,\xi}(t)
\end{equation}
which is the state $\rho_M$ in a one-photon scheme.
This state is the weighted sum of all trajectories which involve a single photon
detection in the interval $[0,t]$, at time $\tau$.
In Appendix~\ref{A1} we show that the state at time $t$, corresponding to a single trajectory involving a photon detection at the instant $\tau$, does not depend on $\tau$.

\section{Efficiency of the protocols} \label{Sec:3}

The formalism developed so far allows us to evaluate the efficiency of the entanglement
protocols described above, in terms of their success probability and fidelity.

\subsection{Single-photon schemes}

We first analyze the schemes based on single-photon detection~\cite{Cabrillo,Browne},
beginning with the set-up without cavities sketched in Fig.~\ref{1:photon}(a), and
considering the situation that during a detection time interval $[0,T_{\rm cw}]$ the two
atoms are weakly driven by a laser which excites spontaneous Raman transitions from
$|e\rangle_j$ to $|g\rangle_j$ via the excited level $|r\rangle_j$. Simultaneously, the
detectors monitor photon emission on the transition $|r\rangle_j\to|g\rangle_j$. The
atoms thus form effective two-level systems, with state $|e\rangle$ decaying to
$|g\rangle$ by spontaneous Raman scattering; the rate of this process will be denoted by
$\Gamma_{\rm eg}$. Then, the master equation governing the atomic dynamics is of the form
given in Eq.~(\ref{masterEq0}), with $A^{(j)}=\ke{g}_j\br{e}$ and with the dissipation
constant $R$ corresponding to $\Gamma_{\rm eg}/2$.

\begin{figure}[!th] \begin{center}
\includegraphics[width=0.5\textwidth]{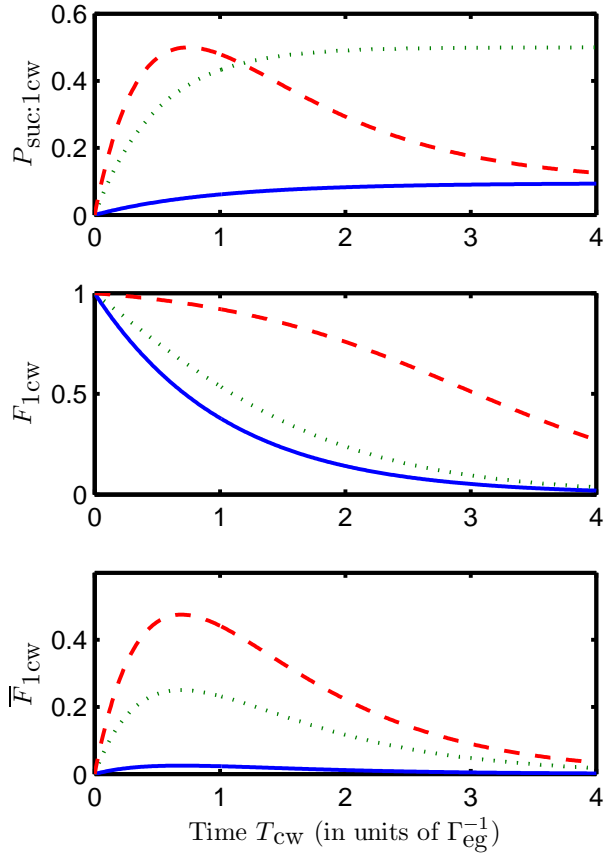}
\caption{Efficiency of the protocol by Cabrillo {\it et al.}~\cite{Cabrillo} when
spontaneous Raman scattering on $|e\rangle \to |g\rangle$ is continuously driven during
the detection time interval $[0,T_{\rm cw}]$. (a) Success probability, (b) conditional
fidelity, and (c) average fidelity, vs.\ $T_{\rm cw}$ and for $\eta=0.05$ (solid line),
$\eta=0.5$ (dotted line), $\eta=0.95$ (dashed line).}\label{figCE}
\end{center} \end{figure}

The probability that a photon is emitted by one atom at any time during $[0,T_{\rm cw}]$
is
\begin{equation}
p_1 = 1-{\rm e}^{-\Gamma_{\rm eg} T_{\rm cw}}. \label{Eq:eps1cw}
\end{equation}
The corresponding success probability, that exactly one photon is recorded at the
detectors, is calculated from Eq.~(\ref{P1}),
\begin{eqnarray}\label{P1ph}
P_{\rm suc:1cw} &=& 2\eta p_1 (1-\eta p_1)~.
\end{eqnarray}
We note that $P_{\rm suc:1cw}\approx 2\eta p_1$ for $p_1 \ll 1$ or $\eta\ll 1$. The
conditional fidelity with which the atoms are in the target state $|\Psi^{\pm}\rangle$
reads
\begin{eqnarray}\label{fidelityCE}
F_{\rm 1cw} =\frac{1-p_1}{1-\eta p_1}~,
\end{eqnarray}
and $F_{\rm 1cw}\approx 1-(1-\eta)p_1$ for $p_1 \ll 1$ or $\eta\ll 1$.
Success probability and conditional fidelity are displayed in Fig.~\ref{figCE}(a)
and~(b), respectively, as functions of $T_{\rm cw}$ and for various values of the
detection efficiency $\eta$. The fidelity is maximum for short times, {\it i.e.}, for
small excitation probability, and then decreases with increasing $T_{\rm cw}$, because
the probability for emission of two photons increases with time. In the regime of small
$p_1$, both success probability and conditional fidelity increase with the detection
efficiency $\eta$. The increase of $F_{\rm 1cw}$ with $\eta$ is due to the fact that
higher $\eta$ makes it more likely that a second emitted photon is also detected, such
that this event would be discarded.

Figure~\ref{figCE}(c) displays the average fidelity $\overline{F}_{\rm 1cw}$ defined in
Eq.~(\ref{F:av}), which for this case reads
\begin{eqnarray}\label{Eq:Fbar1cw}
\overline{F}_{\rm 1cw} &=& 2\eta p_1(1-p_1)~.
\end{eqnarray}
One can see that it increases with the detection efficiency $\eta$, and it always reaches
a maximum at time $T_{\rm cw}=\ln(2)/\Gamma_{\rm eg}$.

The figures give an indication how to choose $p_1$, {\it i.e.}, the detection time
interval $T_{\rm cw}$, at a given detection efficiency $\eta$. Thereby one can optimize
the desired efficiency parameters, putting emphasis on either $F$, $P_{\rm suc}$, or
$\overline{F}$. From Eq.~(\ref{Eq:Fbar1cw}) we find the maximal value for the average
fidelity at $p_1=0.5$. However, with a set-up like the one shown Fig.~\ref{1:photon}(a),
realistic values of $\eta$ are usually in the \%-range or below (see the experimental
considerations in Sec.~\ref{Sec:4}), which results in relatively low conditional fidelity
when $\overline{F}$ is optimal. Larger values of $\eta$ have been included in
Fig.~\ref{figCE} to account for the case that the same scheme of simultaneous excitation
and detection may be applied to a system where the photon collection efficiency is
enhanced by placing a cavity around each atom. In this regime of $\eta$ close to 1, one
may achieve high fidelity at success probabilities close to 0.5, as the dashed curves in
Fig.~\ref{figCE} show.

It should also be mentioned that the assumption of slow Raman excitations, leading to the
scattering probability described by Eq.~(\ref{Eq:eps1cw}), is not strictly necessary. At
stronger excitations, Rabi oscillations on $|e\rangle \leftrightarrow |r\rangle$ appear
and Eq.~(\ref{Eq:eps1cw}) for $p_1$ has to be replaced by one that describes the atomic
dynamics in this regime. Nevertheless, the results for conditional fidelity and success
probability as functions of $p_1$ in Eqs.~(\ref{fidelityCE}) and~(\ref{P1ph}) remain
valid.

\begin{figure*}[!th]
\begin{center}
\includegraphics[width=1.1\textwidth]{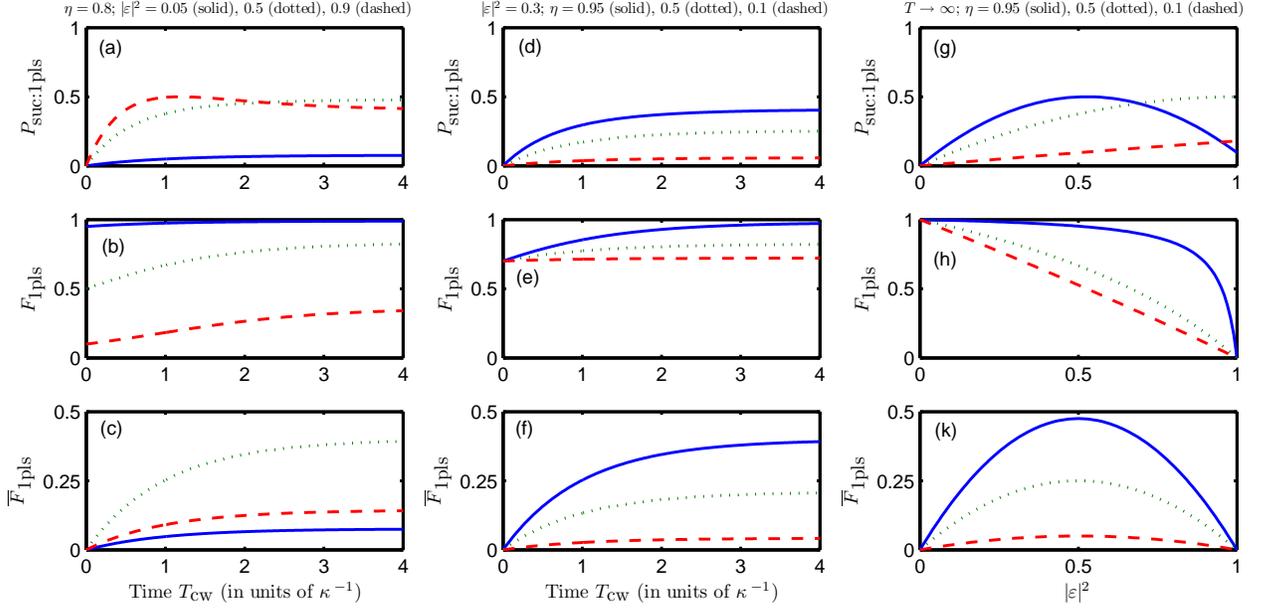}
\caption{Success probability, conditional fidelity, and average fidelity for the one-photon
protocol by Browne {\it et al.}~\protect\cite{Browne}; (a-c) as a function of the
detection time $T$ and for $\eta=0.8$ and $|\varepsilon|^2=0.05,0.5,0.9$ (solid, dotted,
and dashed line, respectively); (d-f) as a function of $T$ and for $|\varepsilon|^2=0.3$
and $\eta=0.95,0.5,0.1$ (solid, dotted, and dashed line, respectively); (g-k) as a
function of $|\varepsilon|^2$ and for $T \to \infty$ and $\eta=0.95,0.5,0.1$ (solid,
dotted, and dashed line, respectively). }\label{figBE}
\end{center}
\end{figure*}

An alternative to this scheme is a protocol where the atoms have been initially prepared
in the state $|\Psi^I\rangle$ by a short laser pulse. The excitation pulse is followed by
a detection time interval $[0,T]$. This version of the protocol has been discussed in
detail in~\cite{Browne} for a set-up including cavities, as shown in
Fig.~\ref{1:photon}(b). Preparation in the state $|\Psi^{I}\rangle$ of
Eq.~(\ref{State:Browne}) is assumed to happen with unit fidelity. The subsequent dynamics
are described by Eqs.~(\ref{masterEq0}) and (\ref{L}), where $R$ is replaced by the
cavity decay constant $\kappa/2$, and $A_{j}$ and $A_{j}\da$ are replaced by the
annihilation and creation operators $a_j$ and $a_j\da$ of photons in cavity $j$. Then the
success probability, {\it i.e.}, the probability to detect exactly one photon in $[0,T]$,
is given by
\begin{eqnarray} \label{P:Success:Bose}
P_{\rm suc:1pls} &=& 2\eta p_{\rm cav} \pt{1-\eta p_{\rm cav}}~,
\end{eqnarray}
where we have set
\begin{eqnarray}
p_{\rm cav}=|\varepsilon|^2(1-{\rm e}^{-\kappa T})~. \label{pcav}
\end{eqnarray}
The conditional fidelity reads
\begin{eqnarray}\label{fidelity10}
F_{\rm 1pls} &=& \frac{1-|\varepsilon|^2}{1-\eta p_{\rm cav}}~,
\end{eqnarray}
and the average fidelity for this scheme is calculated as
\begin{eqnarray}\label{Eq:Fbar1pls}
\overline{F}_{\rm 1pls} &=&  2\eta p_{\rm cav} (1-|\varepsilon|^2)~.
\end{eqnarray}
These quantities are displayed in Fig.~\ref{figBE} as functions of $T$,
$|\varepsilon|^2$, and $\eta$. Like in the previously discussed scheme, large
entanglement ($F$ close to 1) is obtained for small probability that one photon is
created, $|\varepsilon|^2 \ll 1$, see Fig.~\ref{figBE}(b,e,h), because this suppresses
events where two photons are produced in the initial pulse. In contrast to the previous
scheme, however, one observes in Fig.~\ref{figBE}(b) and~(e) that $F$ increases with
increasing $T$. This is easily understood, as the number of photons created in the two
cavities is fixed at $t=0$, and with increasing time the absence of a second detection
enhances the probability that no second photon is present in the cavities. This increase,
on the other hand, depends critically on the detection efficiency, as can be seen in
Fig.~\ref{figBE}(e), because $\eta < 1$ allows for a second photon not being detected
although it was initially created. In the limit $\eta \to 1$, when every created photon
is also detected, high fidelity may be achieved at large success probability for any
value of $|\varepsilon|^2<1$, see Fig.~\ref{figBE}(d,e); the time for reaching a given
large fidelity, nevertheless, becomes longer the closer $|\varepsilon|$ is to unity
(Fig.~\ref{figBE}(b)). The values for $T \to \infty$ are shown in Figs.~\ref{figBE}(g-k).
The average fidelity of this variant of the protocol is shown in Fig.~\ref{figBE}(c,f,k),
and is not substantially improved over the scheme based on continuous excitation and
detection.

\subsection{Two-photon schemes}

We now analyze the efficiency of the two-photon entanglement schemes~\cite{FengDuanSimon},
in a set-up like the one sketched in Fig.~\ref{3scheme}. We assume that the initial state
preparation in $|r_1,r_2\rangle$ is performed with unit fidelity. Out of this state the
atoms decay spontaneously into the state $|\Psi^I\rangle$ of Eq.~(\ref{State:Simon}); the
individual atomic decay rates are taken to be $\Gamma/2$ on both possible transitions,
and photons are collected during the detection time interval $[0,T]$. With proper mode
matching at the beam splitter (but without the need for phase stability), coherence
effects allow for 8 possible two-photon detections (including detection of two photons on
the same detector)~\cite{HongOuMandel}, of which 4 lead to the projection into the target
Bell states. The success probability for this protocol is hence found by multiplying the
probability of two-photon detection, Eq.~(\ref{P2}), by the factor $1/2$, yielding
\begin{eqnarray} \label{P2ph}
P_{\rm suc:2} = \frac{1}{2}\eta^2 p_2^{2}
\end{eqnarray}
where $p_2=1-\exp(-\Gamma T)$ is the probability for each atom to emit a photon at any
instant during the detection time interval $[0,T]$. In this protocol the entanglement
fidelity after detection of two clicks is unity,
\begin{eqnarray}\label{F3scheme}
F_2 = 1~,
\end{eqnarray}
showing that once the successful event occurred, the atoms are found in the corresponding
Bell state independently of $\eta$ and $p_2$. The main limitation to the efficiency of
the two-photon scheme is hence the photon detection probability, which typically gives
small values of $P_{\rm suc:2}$; this might be substantially enhanced in a set-up in
which each atom is coupled to the mode of a resonator. The average fidelity for the
two-photon scheme is equal to the success probability, and reads
\begin{eqnarray}\label{Eq:Fbar2}
\overline{F}_2 = \frac{1}{2}\eta^2 p_2^{2}~.
\end{eqnarray}

\subsection{Single-photon vs two-photon protocols}

We now compare the efficiency of one- and two-photon protocols, focussing on the set-ups where the atoms emit in free space.
First we look at the average fidelity $\overline{F}$: in the one-photon scheme it is maximised to
$\overline{F}_{\rm 1cw}=\eta/2$ by using $p_1=0.5$, while in the two-photon scheme the maximum value
is $\overline{F}_{2}=\eta^2/2$ for $p_2 \to 1$. Thus, in this respect a low detection efficiency
$\eta$ favors the one-photon scheme. A practical assessment also needs to consider the real
time requirements, {\it i.e.}, how many trials the schemes allow to be carried out during some
given experimental time. In this respect, Raman scattering at rate $\Gamma_{\rm eg}$ is
necessarily slower than spontaneous decay at rate $\Gamma$; on the other hand, this is
offset by the fact that $p_2 \to 1$ requires $T \gg 1/\Gamma$, whereas $p_1=0.5$ only
needs time $T_{\rm 1cw} \approx 1/\Gamma_{\rm eg}$.

\begin{figure}[!th] \begin{center}
\includegraphics[width=0.4\textwidth]{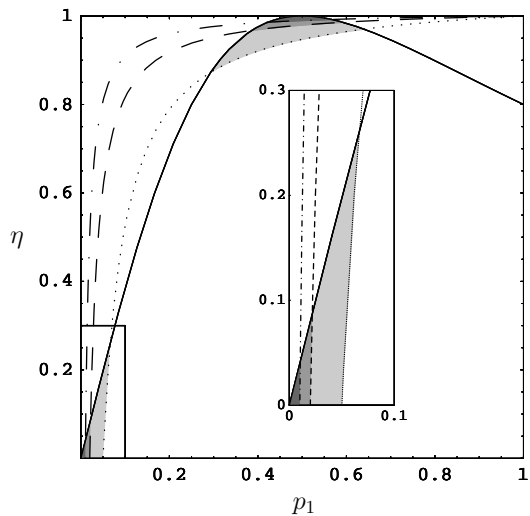}
\caption{Comparison of the efficiency of the one-photon protocol by Cabrillo {\it et
al.}~\cite{Cabrillo} and the two-photon scheme by Simon {\it et al.}~\cite{FengDuanSimon}
when a threshold fidelity $F_{\rm th}$ is set. The horizontal axis is the emission
probability $p_1$ of the one-photon scheme, while $p_2$ has been set to 1. The vertical
axis is the detection efficiency $\eta$ which affects the efficiency of both protocols.
The regions above the dotted, dashed, and dash-dotted lines correspond to $F_{\rm 1cw}>
F_{\rm th}= 0.95, 0.98, 0.99$, respectively. The region below the solid line corresponds
to $P_{\rm suc:1}>P_{\rm suc:2}$. In the shaded parameter regimes both conditions are
satisfied, and under the threshold criterion the one-photon protocol is more efficient than
the two-photon scheme. The inset enlarges the area about the origin.
}\label{1vs2}\end{center}
\end{figure}

If the emphasis is on high fidelity, in the one-photon scheme this can only be reached for
small probability of photon emission, $p_1 \ll 1$. In the two-photon scheme, on the contrary, $F_{\rm 2}=1$ independently of the various
parameters. In order to explore in a practical manner which scheme has larger success probability for
equal fidelity, we fix a threshold value $F_{\rm th}$, such that if $F_{\rm 1cw}>F_{\rm
th}$, we operationally consider the experiment successful. Using Eq.~(\ref{fidelityCE}),
the threshold condition is
\begin{eqnarray}\label{Req:2}
\frac{1-p_1}{1-\eta p_1} > F_{\rm th} ~~\Leftrightarrow~~ p_1 < \frac{1-F_{\rm
th}}{1-\eta F_{\rm th}}~.
\end{eqnarray}
For the corresponding success probabilities we obtain $P_{\rm suc:1cw}>P_{\rm suc:2}$
when
\begin{eqnarray} \label{Req:1}
2\eta p(1-\eta p) > \frac{\eta^2}{2} ~~\Leftrightarrow~~ p-\eta p^2> \frac{\eta}{4}~,
\end{eqnarray}
where we have used Eq.~(\ref{P1ph}), and Eq.~(\ref{P2ph}) in the limit $p_2 \to 1$.
Figure~\ref{1vs2} shows these conditions for various values of $F_{\rm th}$. In the
practically relevant regime $\eta\ll 1$, one finds that fidelity values $F \le 1-\eta/4$
are reached with higher success probability by the one-photon scheme, while for $F >
1-\eta/4$, the two-photon scheme shows higher $P_{\rm suc}$. Like mentioned before, a
full comparison, {\it i.e.}, for real experimental settings, also needs to take into
account the time per execution which the considered protocols require, see
section~\ref{Sec:4}.

\subsection{Purification}

In general, one may consider to increase the success probability $P_{\rm suc:1}$ by
increasing the probability of photon emission $p_1$. Although this occurs at the expense
of fidelity, one can then apply a purification protocol, in order to generate one highly
entangled state out of several states with lower fidelity. A one-photon scheme will then
result more efficient than the corresponding two-photon scheme, if the purification
protocol converges over times shorter than the typical time for a successful two-photon
event.

Following this strategy, we now compare the efficiency of the one-photon protocol
in~\cite{Cabrillo}, to which purification is applied, with the two-photon scheme
in~\cite{FengDuanSimon} under otherwise equal conditions. We require, like before, a
threshold fidelity to be reached in the preparation of a single entangled pair by means
of the creation and subsequent purification of $n$ identical lower-fidelity copies,
$F_{\rm pur}>F_{\rm th}$. Then we compare the total success probability $P_{\rm pur}$
with the success probability of the two-photon scheme, $\eta^2/2$. The total success
probability $P_{\rm pur}$ is given by the product of the success probability for the
preparation of $n$ entangled pairs, $P_{\rm suc:1}/n$,
and the probability for a successful realization of the purification protocol with $n$
prepared entangled pairs, $p_{\rm pur}(n)$:
$$P_{\rm pur}=\frac{1}{n} P_{\rm suc:1} p_{\rm pur}(n)$$.

We evaluate $p_{\rm pur}(n)$ for the purification protocol proposed in~\cite{Deutsch}. It
makes use of $n=2^J$ entangled pairs, with $J$ number of iteration steps of the protocol.
The pairs are divided into $n/2=2^{J-1}$ couples of entangled pairs. Unitary local
operation and local measurements (for which the details are found in~\cite{Deutsch}) are
applied to each couple leading to $n/2$ entangled pairs with fidelity higher than that of
the initial pairs; the remaining $n/2$ pairs are discarded. The protocol is repeated $J$
times until a single highly entangled pair remains. Therefore, if $N_{j-1}$ gives the
probability of having purified one entangled pair at step $j$ of the
protocol~\cite{Deutsch}, then the probability for the successful realization of the
protocol consisting of $J$ steps is
\begin{eqnarray}
p_{\rm pur}(n\equiv2^J)=\prod_{j=1}^J N_{j-1}^{2^{J-j}}~,
\end{eqnarray}
whereby $2^{J-j}$ is the number of couples of entangled pairs at the $j$-th purification
step.

We note that the purification protocol in~\cite{Deutsch} allows for the preparation of
the Bell state $\ke{\Phi^+}$. In order to apply it to the entanglement scheme
in~\cite{Cabrillo}, which allows one to create the Bell state $\ke{\Psi^+}$, one must
perform the unitary transformation $\ke{\Psi^+}\to\ke{\Phi^+}$ before the application of
the purification~\cite{Footnote:Purification}. Then, at the end of the protocol one
inverts the transformation.

\begin{figure}[!th] \begin{center}
\includegraphics[width=0.45\textwidth]{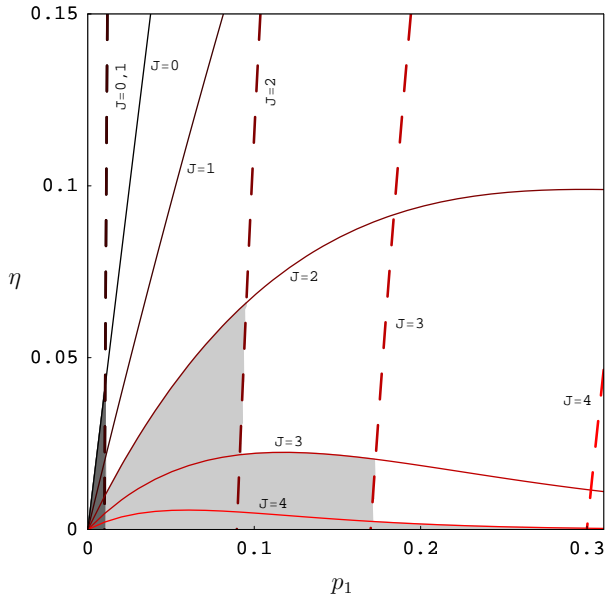}
\caption{Efficiency with purification. The area on the left of each dashed curve
corresponds to the region of parameters $p_1$ and $\eta$ where $F_{\rm pur}>F_{\rm
th}=0.99$ when the state is prepared with $J=0,1,2,3,4$ iteration steps of the
purification protocol (note that $J=0$ corresponds to the result obtained without
purification, see Fig.~\ref{1vs2}). Similarly, the areas below the solid lines correspond
to the parameters where $P_{\rm pur}>P_{\rm suc:2}$ after $J$ purification steps.
}\label{1vs2pur}\end{center}
\end{figure}

In Fig.~\ref{1vs2pur} we compare the efficiency of the purified single-photon scheme with
the efficiency of the corresponding two-photon scheme. The threshold fidelity is set to
$F_{\rm th}=0.99$. The dark-gray shaded area displays the parameter region where the
single-photon scheme without purification is more efficient than the two-photon scheme,
{\it i.e.}, where $F_{\rm 1cw}>0.99$ and $P_{\rm suc:1}>P_{\rm suc:2}$, see also
Fig.~\ref{1vs2}. The light-gray shaded area, on the other hand, displays the parameter
regime where the purified single-photon scheme is more efficient than the two-photon
protocol, $F_{\rm pur}>0.99$ and $P_{\rm pur}>P_{\rm suc:2}$, showing that this is also
the case for larger values of $p_1$ and of $\eta$. As $p_1$ is increased, one needs a
larger number of iteration steps in order to reach fidelities $F_{\rm pur}\geq F_{\rm
th}$, and the corresponding success probability $P_{\rm pur}$ decreases.

\section{Experimental Considerations}
\label{Sec:4}

To estimate what fidelities and success probabilities can be
realistically achieved in experimental implementations of the
proposals discussed mentioned, we will consider typical parameters
of current ion trapping experiments.

Benchmark values for the single-photon scheme proposed by Cabrillo
{\it et al.}~\cite{Cabrillo} are derived for the ion trapping
experiment with $^{40}$Ca$^+$ ions which is being operated by the
authors~\cite{Gerber}. Here, the three states $|e\rangle$,
$|g\rangle$, and $|r\rangle$, are identified with the levels
S$_{1/2}(m=-\frac{1}{2})$, S$_{1/2}(m=+\frac{1}{2})$, and
P$_{1/2}(m=-\frac{1}{2})$, respectively. The transitions
connecting the P-state with the two S-states are distinguished by
their polarization, which allows for preparing the initial state
$|e\rangle$ by optical pumping, as well as for identifying a
photon emitted on $|r\rangle \to |g\rangle$. To reach high entanglement fidelities, the probability of exciting
both atoms rather than just one has to be kept small, which is the
case for low emission probability of a photon on $|r\rangle \to
|g\rangle$. The probability for this spontaneous Raman process to
happen during the detection time interval $[0,T_{\rm cw}]$ is
given by
\begin{eqnarray}\label{p1cabrillo}
p_1 = \alpha_{\rm rg} (1-e^{-\gamma'T_{\rm cw}})~,
\end{eqnarray}
where $\gamma'$ is the effective rate at which state $|e\rangle$
is emptied by the laser excitation, and $\alpha_{\rm rg}$ is the
branching ratio for an atom to decay into $|g\rangle$ (see
Appendix~\ref{Appendix:exp}). Any desired value of the emission
probability $p_1$ is easily achieved in the experiment through
control of the Rabi frequency, {\it i.e.}, by adjusting the laser
intensity. The detection time interval in which the laser is
applied and the detectors are gated can be as short as $12.5$~ns
for the considered apparatus. Setting the Rabi frequency to
$\Omega_{\rm er} = 2 \pi \times 5.4$~MHz one achieves an emission
probability of $p_1= 0.15$ for $T_{\rm cw}\approx 200$~ns.
The detection efficiency of the setup is $\eta = \eta_d \chi
\approx 0.5\%$, estimated from the detector quantum efficiency,
$\eta_d \approx 0.25$, for photons emitted on the $P_{1/2}\to
S_{1/2}$ transition (397~nm), and from the photon collection
efficiency $\chi = (\Delta \Omega / 4 \pi) L \approx 2\%$. The
latter is determined by the solid angle within which photons are
collected with a high numerical aperture lens ($NA=0.4$) and the
transmission losses in optical elements and fiber coupling,
$L\approx 0.5$.
For these values of detection efficiency and emission probability
one expects to obtain a conditional fidelity $F_{\rm 1cw}=85\%$ at
a success probability $P_{\rm suc:1cw}=1.5\times10^{-3}$, leading
to an average fidelity $\overline{F}_{\rm 1cw}=1.3\times10^{-3}$,
see Eqs.~(\ref{P1ph}, \ref{fidelityCE}, \ref{Eq:Fbar1cw}). It
seems feasible to run the protocol at a rate of $10^5$
experimental sequences per second, which provides sufficient time
for state preparation, detection and cooling. Hence, about 150
entangled pairs per second of the above fidelity may be generated.
Since the scheme relies on single photon interference for the
creation of entanglement, interferometric stability of the setup
is required, which is experimentally
challenging.

One may hope to obtain better values when increasing the detection
efficiency by placing the atoms in cavities and preparing
entangled atom-cavity states by a short initial laser pulse, as
described in section~\ref{Sec:1}. Here we consider such a system
on the basis of an experimental implementation with a single
trapped calcium ion coupled to a high-finesse
cavity~\cite{Carlos}. The cavity, in this case, couples strongly to the ${\rm P}_{1/2}
\to {\rm D}_{3/2}$ transition, while the ion is laser-excited on
the ${\rm S}_{1/2} \to {\rm P}_{1/2}$ transition. For the given
setup the decay constant of the cavity field was determined to be
$\kappa/2 = 2\pi \times 54$~kHz and photons were detected at a
rate of $33$~kHz for a mean intra-cavity photon number of $\langle
n \rangle = 1$, from which we infer the detection efficiency $\eta
= 0.31$.
We identify the S$_{1/2}$, P$_{1/2}$, and D$_{3/2}$ states with
the three levels of the $\Lambda-$system, $|e\rangle$, $|r\rangle$
and $|g\rangle$, respectively. Initially the atoms are assumed to
be prepared in the state $|\Psi^I\rangle$ of
Eq.~(\ref{State:Browne}) by a laser pulse, which drives a
cavity-assisted coherent Raman transition on $|e\rangle \to
|g\rangle$. The Rabi frequency and the detunings of laser and
cavity are chosen such that the probability to make the Raman
transition is $|\varepsilon|^2 = 0.15$, comparable to the previous
scenario.
For a given initial state and a fixed value of the cavity decay
rate, the probability that a photon is emitted by the atom-cavity
system, $p_{\rm cav}$ of Eq.~(\ref{pcav}), is determined only by
the detection time interval, $T$. Since the conditional fidelity,
Eq.~(\ref{fidelity10}), for this set of parameters varies only in
the range of $85\%$ to $89\%$ with detection time (see also
Fig.\ref{figBE}(e)), one may choose to optimize the number of
detection events per time interval, which is the experimental rate
of scheme executions times the success probability of
Eq.~(\ref{P:Success:Bose}). The optimum value of about $2000$
events per second is found for detection time intervals of
approximately 10~$\mu$s, which results in $p_{\rm cav} = 0.1$.
Allowing for additional 20~$\mu$s for cooling and for the
preparation of the initial atom-cavity state, we obtain a rate of
$3.3\times 10^4$ experimental sequences per second. From these
values we calculate the expected success probability, $P_{\rm
suc:1pls}=6.0\times 10^{-2}$, the fidelity, $F_{\rm 1pls}=0.88$
and the average fidelity, $\overline{F}_{\rm
1pls}=5.3\times10^{-2}$. As expected, the higher detection
efficiency due to the increased collection efficiency of
fluorescence photons emitted from an atom coupled to a cavity,
leads to higher success probabilities at similar values of the
conditional fidelities as compared to the free-space scheme
considered above. However, the experimental implementation of a
single ion coupled to a high-finesse cavity
is considerably more demanding than the free-space case.

Creation of entanglement through a two-photon
scheme~\cite{FengDuanSimon} has been experimentally realized in
Ref.~\cite{Maunz,Matsukevich}. Two remotely located trapped
$^{171}$Yb$^+$ ions were entangled with fidelity $F_2=81\%$, and a
Bell inequality violation by more than three standard deviations
was reported. We base our evaluation of the efficiency of the
scheme on the experiment of Ref.~\cite{Matsukevich}.
In this work, each ion is initially prepared with near unit
efficiency in its $^2$P$_{1/2}, |F=1, m_F=0\rangle$ state from
where it decays with equal probability to either of three
$^2$S$_{1/2}$ states, emitting a $369.5$~nm photon. Observing
along the quantization axis one detects $\sigma^-$ and
$\sigma^+$-polarized photons from the decay to the $|1, 1\rangle$
and $|1, -1\rangle$ states, respectively, while $\pi$-polarized
photons emitted in the decay to the $|0, 0\rangle$ state do not
propagate along this direction. For detection time intervals
significantly longer than the excited state lifetime ($\simeq
8$~ns), the emission probability consequently reads $p_2 = 2/3$.
Just as in the theoretical proposal, the polarization of each
emitted photon is then entangled with the state of its respective
ion. The photons are collected with high numerical aperture lenses
and coupled into single mode fibers before they are overlapped on
a free-space beam splitter. Interference of the single photon
wavepackets and coincidence detection of one photon at each output
port of the beam splitter projects the two ions into one of the
four Bell states. Considering the quantum efficiency of the
detectors, transmission through optical elements, fiber coupling,
and the solid angle of photon collection, the detection efficiency
was $\eta = 6.7\times 10^{-4}$.
At a typical experiment repetition rate of $5.2\times10^{5}$ per
second, the generation of one entangled atom pair of the above
fidelity every $39$ seconds was reported, from which we infer a
success probability of $P_{\rm suc:2}=\frac{1}{4}\eta^2
p_2^2=4.9\times 10^{-8}$. The factor $1/4$ accounts for the fact
that one out of all four Bell states was detected. With a
conditional fidelity $F_2$ of $81\%$, estimated from the density
matrix obtained by state tomography, the average fidelity reads
$\overline{F}_2=4\times10^{-8}.$ The measured fidelity was limited
by experimental imperfections in the generation of ion-photon
entanglement and imperfect interference contrast of the photon
modes at the beam splitter.
It has to be pointed out that this scheme has the advantage to
rely on two-photon interference, which significantly relaxes the
conditions on the experimental setup as compared to the
single-photon schemes, which require interferometric stability.
This has certainly contributed to making the two-photon protocol
the first one with which distant entanglement of single atoms was
demonstrated.

\section{Conclusions} \label{Sec:5}

Based on recent proposals and experimental progress, we have studied the efficiency of
protocols for entangling distant atoms by projective measurement of emitted photons,
focussing on the role of the photon detection efficiency. We distinguish schemes based on
one-photon and two-photon detection. For their comparison we have calculated, using an
unravelling of the master equation, their success probability and conditional fidelity.
We conclude that for low detection efficiency, which is typical for current experiments,
and when the fidelity is assumed to be similar in both methods, protocols based on the
detection of a single photon exhibit larger success probability. This includes the
possibility of applying state purification. We calculated the efficiency criteria for
concrete situations encountered in recent or ongoing experiments. Beyond these criteria,
the decision which method to choose in a given experimental setting will also have to
take into account that the experimental difficulty of the schemes may be significantly
different (the only experimental result so far is based on two-photon
detection~\cite{Maunz}), and that there may be a detrimental effect of dark counts (see Ref.~\cite{MonroeJOSAB}). With marginal differences, the results of
our analysis can be extended to the efficiency of teleportation protocols based on
projective measurement~\cite{Bose,FengDuanSimon}.

While the calculation of the efficiency criteria may vary with the physical system, we
believe that these concepts are generally applicable to all systems that may be considered
for the creation of distant entanglement, including atomic-ensemble, photonic, and solid
state implementations. In all such systems the detection efficiency will have a similar,
important role for the use of the entanglement as a resource in quantum technologies.

\acknowledgments

We acknowledge support by the European Commission (EMALI, MRTN-CT-2006-035369; SCALA,
Contract No.\ 015714) and by the Spanish Ministerio de Educaci\'on y Ciencia (QOIT,
Consolider-Ingenio 2010 CSD2006-00019; QLIQS, FIS2005-08257; QNLP, FIS2007-66944;
Ramon-y-Cajal). G.A.O.-R. thanks CONICyT for scholarship support. C.S. acknowledges
support by the Commission for Universities and Research of the Department of Innovation,
Universities and Enterprises of the Catalonian Government and the European Social Fund.

\begin{appendix}

\section{}
\label{A1}

In this appendix we show that, when a single photo-detection event occurs at the instant $\tau$ in the time interval $[0,t]$, the corresponding state of the system at time $t$ is the same independently of the specific instant $\tau$ in which the
photon was measured. This implies that the state $\rho_{+,\xi}(t)$ in Eq.~(\ref{rho+xi}) satisfies the equation $\rho_{+,\xi}(t)=\bar\rho_{+,\xi}(t,\tau)$, where 
\begin{equation}
\label{rho:tau}
\bar\rho_{+,\xi}(t,\tau)=U(t-\tau){\cal C}_{+,\xi}U(\tau)\rho(0)/{\mathcal N}_{+,\xi}(t,\tau)
\end{equation}
is the state corresponding to a single trajectory when a single photon is measured at time $\tau$ in the detection interval $[0, t]$, and 
${\mathcal N}_{+,\xi}(t,\tau)={\rm Tr}\{U(t-\tau){\cal C}_{+,\xi}U(\tau)\rho(0)\}$.

For simplicity, we just restrict to single-photon schemes and therefore we do not use the index $\xi$ for the field polarization in the operators $A_\xi\al{j}$. In our demonstration we will use the following relations.
Since in the considered setup each atom can emit at most a single photon then 
\begin{eqnarray}\label{AA0}
A\al{j}A\al{j}\rho(0)=\rho(0) {A\al{j}}\da {A\al{j}}\da=0\,.
\end{eqnarray}
This relation implies
\begin{eqnarray}\label{AA1}
&&{\rm e}^{-R t\ {A\al{j}}\da A\al{j}}A\al{j}\rho(0)=A\al{j}\rho(0) \\
&&A\al{j}{\rm e}^{-R t\ {A\al{j}}\da A\al{j}}\rho(0)={\rm e}^{-R t}A\al{j}\rho(0) \label{AA2}
\end{eqnarray}
Using relation~(\ref{AA0}) we rewrite the evolution operator $U(t)$ in Eq.~(\ref{rho:tau}) as
\begin{eqnarray}\label{U0U1}
U(t)\rho=\pq{U_0(t)+(1-\eta)U_1(t)}\rho\,,
\end{eqnarray}
with
\begin{eqnarray}
&&U_0(t)\rho= {\rm e}^{-R t\ \sum_j {A\al{j}}\da A\al{j}}\rho {\rm e}^{-R t\ \sum_j {A\al{j}}\da A\al{j}}\,,\\
&&U_1(t)\rho=\int_0^t dt' U_0(t-t')JU_0(t')\rho \nonumber\,.
\end{eqnarray}
We use Eqs.~(\ref{AA1})-(\ref{AA2}) in Eq.~(\ref{U0U1}), and consider the term $U(t-\tau){\cal C}_+ U(\tau)\rho(0)$ in Eq.~(\ref{rho:tau}). After some algebra, we get
\begin{eqnarray}\label{UCUrho}
&&U(t-\tau){\cal C}_+ U(\tau)\rho(0)={\rm e}^{-2R\tau}\lpq{\chi(t)\rho(0)\chi\da(t) }\nonumber\\
&&+\rpq{ 2\pt{1-{\rm e}^{-2Rt}}  A\al{1}A\al{2}\rho(0){A\al{1}}\da {A\al{2}}\da }\,,
\end{eqnarray}
with
\begin{eqnarray}
\chi(t)
=A\al{1}{\rm e}^{-R t\ {A\al{2}}\da A\al{2}}+A\al{2}{\rm e}^{-R t\ {A\al{1}}\da A\al{1}}\,.
\end{eqnarray}
Substituting Eq.~(\ref{UCUrho}) in Eq.~(\ref{rho:tau}) we obtain $\bar\rho_{+,\xi}(t,\tau)=\rho_{+,\xi}(t)$.

\section{}
\label{Appendix:exp}

The effective rate $\gamma'$ at which population is removed from the initial state
$|e\rangle$ is given by the rates at which population is transferred either to the state
$|g\rangle$ or out of the considered three-level system to the D$_{3/2}$-state,
$|D\rangle$, to which the P-level can also decay. It reads
\begin{eqnarray}
\gamma'=\frac{\Omega_{\rm er}^2}{\Gamma_r^2}(\Gamma_{\rm rg}+\Gamma_{\rm rD})\,,
\end{eqnarray}
where $\Omega_{\rm er}$ is the Rabi frequency of the laser driving the transition
$|e\rangle\to |r\rangle$, $\Gamma_{\rm r}=21.7$~MHz is the total decay rate of the
excited state, and $\Gamma_{\rm rg}= 13.3$~MHz and $\Gamma_{rD}= 1.7$~MHz are the decay
rates from state $|r\rangle$ into states $|g\rangle$ and $|D\rangle$, respectively.

The branching ratio $\alpha_{\rm rg}$ is, correspondingly,
\begin{eqnarray}
\alpha_{\rm rg} = \frac{\Gamma_{\rm rg}}{\Gamma_{\rm rD}+\Gamma_{\rm rg}} = 89\%~.
\end{eqnarray}
These modifications to the formalism in Sec.~\ref{Sec:2} must be applied because of the
presence of the additional decay channel to state $|D\rangle$.

\end{appendix}


\begin{thebibliography}{99}

\bibitem{ZollerRoadmap}
P. Zoller {\it et al.}, {\it Quantum information processing and communication}, Eur.
Phys. J. D {\bf 36}, 203 (2005).

\bibitem{Briegel98}
H. J. Briegel, W. D\"{u}r, J. I. Cirac, P. Zoller, Phys. Rev. Lett. {\bf 81}, 5932
(1998).

\bibitem{Cirac97}
J. I. Cirac, P. Zoller, H. J. Kimble, and H. Mabuchi, Phys. Rev. Lett. \textbf{78}, 3221 (1997).

\bibitem{Kraus04}
B. Kraus and J. I. Cirac, Phys. Rev. Lett. \textbf{92}, 013602 (2004).

\bibitem{Cabrillo}
C. Cabrillo, J. I. Cirac, P. Garc{\'i}a-Fern{\'a}ndez, and P. Zoller, Phys. Rev. A {\bf
59}, 1025 (1999).

\bibitem{Browne}
D. E. Browne, M. B. Plenio, and S. F. Huelga, Phys. Rev. Lett. {\bf 91}, 067901 (2003).


\bibitem{FengDuanSimon}
X.-L. Feng, Z.-M. Zhang, X.-D. Li, S.-Q. Gong, Z.-Z. Xu, Phys. Rev. Lett. {\bf 90},
217902 (2003); L.-M. Duan, H. J. Kimble, Phys. Rev. Lett. {\bf 90}, 253601 (2003); C.
Simon, W. T. M. Irvine, Phys. Rev. Lett. {\bf 91}, 110405 (2003).

\bibitem{Eschner2001}
J. Eschner, Ch. Raab, F. Schmidt-Kaler and R.Blatt, Nature {\bf
413}, 495 (2001).

\bibitem{Eichmann1993}
U. Eichmann, J. C. Bergquist, J. J. Bollinger, J. M. Gilligan, W. M. Itano, D. J.
Wineland, and M. G. Raizen, Phys. Rev. Lett {\bf 70}, 2359 (1993);

\bibitem{Walther04}
M. Keller, B. Lange, K. Hayasaka, W. Lange, H. Walther, Nature {\bf 431}, 1075 (2004).

\bibitem{Kimble04}
J. McKeever, A. Boca, A. D. Boozer, R. Miller, J. R. Buck, A. Kuzmich, H. J. Kimble,
Science {\bf 303}, 1992 (2004).

\bibitem{Grangier05}
B. Darquie, M. P. A. Jones, J. Dingjan, J. Beugnon, S. Bergamini, Y. Sortais, G. Messin,
A. Browaeys, P. Grangier, Science {\bf 309}, 454 (2005).

\bibitem{Grangier06}
A. Ourjoumtsev, R. Tualle-Brouri, J. Laurat, P. Grangier, Science {\bf 312}, 83 (2006).

\bibitem{Kuhn2}
M. Hijlkema, B. Weber, H.  P. Specht, S. C. Webster, A. Kuhn, and G. Rempe, Nature
Physics {\bf 3}, 253 (2007).

\bibitem{Rempe07}
T. Wilk, S. C. Webster, A. Kuhn, and G. Rempe, Science {\bf 317},
488 (2007).

\bibitem{Kimble08}
B. Dayan, A. S. Parkins, T. Aoki, E. P. Ostby, K. J. Vahala, and H. J. Kimble, Science
{\bf 319}, 1062 (2008).


\bibitem{Monroe04}
B. B. Blinov, D. L. Moehring, L.-M. Duan, C. Monroe, Nature {\bf 428}, 153-157 (2004).

\bibitem{Weinfurter06}
J. Volz, M. Weber, D. Schlenk, W. Rosenfeld, J. Vrana, K. Saucke, C. Kurtsiefer, H.
Weinfurter, Phys. Rev. Lett. {\bf 96}, 030404 (2006).

\bibitem{Maunz}
D. L. Moehring, P. Maunz, S. Olmschenk, K. C. Younge, D. N.
Matsukevich, L.-M. Duan, and C. Monroe, Nature {\bf 449}, 68
(2007).

\bibitem{Kimble-Ensemble}
C. W. Chou, H. de Riedmatten, D. Felinto, S. V. Polyakov, S. J.
van Enk, and H. J. Kimble, Nature {\bf 438}, 828 (2005).


\bibitem{Kuzmich-Ensemble}
D. N. Matsukevich, T. Chaneli{\`e}re, S. D. Jenkins, S.-Y. Lan, T. A.
B. Kennedy, and A. Kuzmich, Phys. Rev. Lett. {\bf 96}, 030405
(2006).

\bibitem{Polzik-Teleportation}
J.F. Sherson, H. Krauter, R.K. Olsson, B. Julsgaard, K. Hammerer, I. Cirac, and E.S. Polzik1, Nature {\bf 443}, 557 (2006).

\bibitem{Schmiedmayer}
Y.-A. Chen, S. Chen, Z.-S. Yuan, B. Zhao, C.-S. Chuu, J. Schmiedmayer, J.-W. Pan,
Nature Physics {\bf 4}, 103 (2008).

\bibitem{Lukin}
L. Childress, J. M. Taylor, A. S. S{\o}rensen, and M. D. Lukin, Phys. Rev. A {\bf 72},
052330 (2005).

\bibitem{LamataCirac}
L. Lamata, J.J. Garcia-Ripoll, and J.I. Cirac, Phys. Rev. Lett. {\bf 98}, 010502 (2007).


\bibitem{Solano}
C. Thiel, J. von Zanthier, T. Bastin, E. Solano, and G. S. Agarwal,
Phys. Rev. Lett. {\bf 99}, 193602 (2007).

\bibitem{Deutsch}
D. Deutsch, A. Ekert, R. Jozsa, C. Macchiavello, S. Popescu, A. Sanpera, Phys. Rev. Lett.
{\bf 77}, 2818 (1996).

\bibitem{MonroeJOSAB}
D. L. Moehring, M. J. Madsen, K. C. Younge, R. N. Kohn, Jr., P. Maunz, L.-M. Duan, C.
Monroe, and B. Blinov, J. Opt. Soc. Am. B {\bf 24}, 300 (2007).

\bibitem{Bose}
S. Bose, P. L. Knight, M. B. Plenio, V. Vedral, Phys. Rev. Lett. {\bf 83}, 5158 (1999).

\bibitem{Footnote} For simplicity, we neglect the dependence of
the probability amplitude on the mode (direction of space) in
which the photon is emitted.

\bibitem{Footnote:Mode}
If one instead uses a plane-wave decomposition, the
direction of the emitted photons is distributed over the whole
solid angle according to the corresponding dipole pattern of
emission.

\bibitem{HongOuMandel}
C. K. Hong, Z. Y. Ou, and L. Mandel, Phys. Rev. Lett. {\bf 59}, 2044 (1987).


\bibitem{Englert} For a review, see B.-G. Englert and G. Morigi, in
{\it Coherent Evolution in Noisy Environments}, Lecture Notes in Physics {\bf 611}, p.
55, ed. by A. Buchleitner, K. Hornberger (Springer Berlin, Heidelberg, New York 2002),
and references therein.

\bibitem{Carmichael}
H. J. Carmichael, {\it An open system approach to quantum optics}, Springer Verlag
(Berlin-Heidelberg-New York, 1993).

\bibitem{Dum}
R. Dum, P. Zoller, and H. Ritsch, Phys. Rev. A {\bf 45}, 4879 (1992).

\bibitem{MilburnWalls}
D. F. Walls and G. J. Milburn, {\it Quantum Optics}, Springer Verlag
(Berlin-Heidelberg-New York, 1994).

\bibitem{Footnote:Inputnoise}
The input noise terms give rise to the dark counts discussed
in~\protect\cite{MonroeJOSAB}. They can be discarded in our formalism as we assume that
the external electromagnetic field is in the vacuum. Hence, their contribution to the
photo-detection signal vanishes.


\bibitem{Footnote:Purification} It can be realized for example by
the local unitary operation
$\sigma_x\al{1}\ke{\psi^+}=\ke{\phi^+}$ where
$\sigma_x\al{1}=\ke{e}_1\br{g}+\ke{g}_1\br{e}$ operates only on
the first atom.

\bibitem{Gerber}
See
M. Hennrich, {\it et al.}, in preparation; see also
http://www.icfo.es/groups/eschner. 


\bibitem{Carlos}
F. Dubin, private communication.

\bibitem{Matsukevich}
D. N. Matsukevich, P. Maunz, D. L. Moehring, S. Olmschenk, and C. Monroe, Phys. Rev.
Lett. {\bf 100}, 150404 (2008).


\end{thebibliography}
\end{document}